\documentclass[showkeys, showpacs, preprintnumbers, amsmath, amssymb, floatfix, twocolumn]{revtex4-1}
\usepackage{graphicx}
\usepackage{dcolumn}
\usepackage{bm}

\begin{document}

\title{Thermoelectric Power of the YbT$_{2}$Zn$_{20}$ (T = Fe, Ru, Os, Ir, Rh, and Co) Heavy Fermions}
\author{E. D. Mun\footnote{Current address : National High Magnetic Field Laboratory, Los Alamos National Laboratory, Los Alamos, NM 87545, USA}, S. Jia\footnote{Current address : Department of Chemistry, Princeton University, Princeton, New Jersey 08544, USA}, S. L. Bud'ko, P. C. Canfield}
\affiliation{Ames Laboratory US DOE and Department of Physics and Astronomy, Iowa State University, Ames, IA 50011, USA}%

\date{\today}

\begin{abstract}
The thermoelectric power, $S(T)$, of the heavy fermions YbT$_{2}$Zn$_{20}$ ($T$ = Fe, Ru, Os, Ir, Rh, and Co) has been measured to shed further light on their strong electronic
correlations. A large, negative, local minimum in $S(T)$ with approximately -70\,$\mu$V/K is found for all compounds. From the observed local minimum, the energy scales associated
with both the Kondo temperature and the crystalline electric field splitting are deduced and compared to previous specific heat measurements. At low temperatures, a highly enhanced
$S(T)/T$ value is observed for all members, although $S(T)$ does show a deviation from a purely linear temperature dependence, $S(T)$ = $\alpha T$, for T $\neq$ Fe members. In the
zero temperature limit, estimated by a simple linear extrapolation, the enhanced $S(T)/T$ value strongly correlates with the electronic specific heat coefficient, $C(T)/T$.
\end{abstract}


\maketitle

\section{Introduction}
In a heavy fermion (HF) Kondo lattice system, the ground state is a Fermi-liquid (FL) state formed out of Landau quasi-particles. In Ce-, Yb-, and U-based intermetallic systems the
conduction electrons compensate, or screen, the localized moments of the $f$-electrons where localized electrons together with their screening cloud form quasi-particles. These
quasi-particles have heavy masses, reflected in an enhanced value of the Sommerfeld coefficient, $\gamma$ = $C(T)/T|_{T\rightarrow 0}$, at low temperatures \cite{Hewson1993}.

For Yb-based HF systems, the electrical resistivity and thermoelectric power (TEP) reveal complex temperature dependencies with a local extrema. In general, these extrema are related
to Kondo scattering associated with the ground state and excited states of the CEF energy levels \cite{Bhattacharjee1976, Lassailly1985, Maekawa1986}. The characteristic temperature of
the local maximum shown in $\rho(T)$ and the local minimum developed in $S(T)$ allow for an estimate of the Kondo temperature, $T_{K}$, and the crystalline electric field (CEF)
splitting, $\Delta/k_{B}$, as relevant energy scales in Yb-based HF systems.

The FL state in HF Kondo lattice systems shows strong correlations among physical quantities. One such correlation is the Kadowaki-Woods (K-W) ratio, a relation between the electrical
resistivity ($\rho(T) - \rho_{0}$\,=\,$AT^{2}$) and specific heat ($C(T)$\,=\,$\gamma T$), given by what was originally thought to be an universal ratio $A/\gamma^{2}$ =
1.0$\times$10$^{-5}$ $\mu\Omega$cm/(mJ/mol$\cdot$K)$^{2}$ \cite{Kadowaki1986, Miyake1989}. Recently, systematic deviations of the K-W ratio in many HF systems (especially for Yb-based
compounds) have been explained by Tsujii $et$ $al$., taking into account the ground state degeneracy ($N$\,=\,2$j$+1) \cite{Tsujii2003, Kontani2004, Tsujii2005}. A FL state is also
characterized by the Wilson ratio ($R_{W}$) which links $\gamma$ to the Pauli susceptibility $\chi(0)$ \cite{Weigman1983, Auerbach1986, Lee1986}, which is given by $R_{W}$ =
$\pi^{2}k^{2}_{B}\chi(0)/(j(j+1)g^{2}\mu^{2}_{B}\gamma^{2})$, where $k_{B}$, $g$, and $\mu_{B}$ are the Boltzman constant, Lande's factor, and Bohr magneton, respectively
\cite{Hewson1993}. In addition to the $R_{W}$ and the K-W ratio, the zero temperature limit of the TEP divided by temperature, $S(T)/T$\,=\,$\alpha$, for several correlated materials
has shown a strong correlation with $\gamma$ via the dimensionless ratio $q$\,=\,$N_{A}eS/\gamma T$ = $N_{A}e\alpha/\gamma$ \cite{Behnia2004}, where $N_{A}$ is the Avogadro number and
$e$ is the carrier charge. It is very important to test the universality of the K-W ratio, Wilson's number, and q ratio for a number of isostructural materials; the YbT$_{2}$Zn$_{20}$
compounds offer six isostructural materials where the local environment of the hybridizing Yb-ion is identical (nearest and next nearest neighbors are all Zn).

In this paper, TEP measurements on YbT$_{2}$Zn$_{20}$ (T = Fe, Ru, Os, Ir, Rh, and Co) are presented as functions of temperature and magnetic field to study their temperature and
magnetic field dependence and to evaluate the correlation between specific heat and TEP in the zero temperature limit. These compounds crystallize in the cubic
CeCr$_{2}$Al$_{20}$-type structure ($F\,d\,\overline{3}\,m$, No.227) \cite{Thiede1998} and have been reported to be HF metals with no long range order down to 20\,mK
\cite{Torikachvili2007}. In the FL regime it has been shown that the $R_{W}$ and K-W ratios in this family follow the theoretical predictions with different ground state degeneracies.
The TEP data of YT$_{2}$Zn$_{20}$ (T = Fe, Co) are also presented for comparison. YFe$_{2}$Zn$_{20}$ is an example of a nearly ferromagnetic Fermi liquids (NFFL) with a highly
enhanced magnetic susceptibility value at low temperatures \cite{Jia2007}, whereas YCo$_{2}$Zn$_{20}$ shows un-enhanced Pauli paramagnetic, metallic behavior.

\section{Experimental}
Single crystals of YbT$_{2}$Zn$_{20}$ (T = Fe, Ru, Os, Ir, Rh, and Co) and isostructural YT$_{2}$Zn$_{20}$ (T = Fe and Co) were grown out of excess Zn \cite{Torikachvili2007, Jia2007}
using standard solution growth techniques \cite{Canfield1992, Canfield2010}. The TEP was measured using a dc, alternating heating, technique that utilizes two heaters and two
thermometers \cite{Mun2010}. A Quantum Design Physical Property Measurement System provided the temperature (from 2 to 300\,K) and magnetic field (up to 140\,kOe) environment. For T =
Fe, Rh, and Co, zero-field TEP measurements were extended down to 0.4\,K, measured using the same technique \cite{Mun2010}, in a CRYO Industries of America, $^3$He system. The heat
current was generated in the (111)-plane of the samples ($\Delta$\,$T$\,$\parallel$\,(111)) and the magnetic field was applied along the [111]-direction maintaining a transverse
configuration, $\Delta T$\,$\perp$\,\textbf{H}.

\begin{figure}
\centering
\includegraphics[width=1\linewidth]{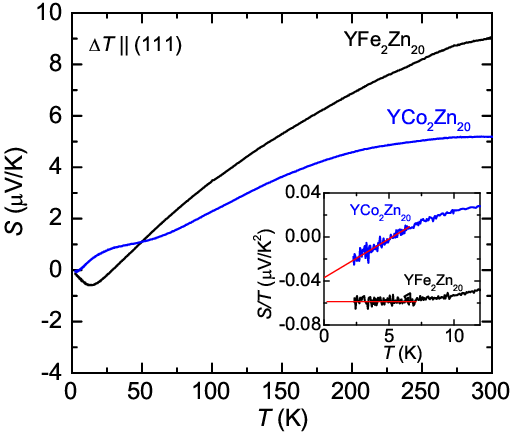}
\caption{Temperature-dependent thermoelectric power, $S(T)$, of YFe$_{2}$Zn$_{20}$ and YCo$_{2}$Zn$_{20}$ for $\Delta T$\,$\parallel$\,(111). Inset: $S(T)/T$ vs. $T$ below 12\,K. Solid
lines are guide to eye.}
\label{Zn20YST}%
\end{figure}%

\section{Results}

Figure \ref{Zn20YST} shows the TEP data for YFe$_{2}$Zn$_{20}$ and YCo$_{2}$Zn$_{20}$. The temperature-dependent TEP, $S(T)$, of these compounds is similar to normal metallic systems.
At 300\,K, $S(T)$ of both compounds is positive and has an absolute value of $\simeq$\,9\,$\mu$V/K for YFe$_{2}$Zn$_{20}$ and $\simeq$\,5\,$\mu$V/K for YCo$_{2}$Zn$_{20}$, and then
decrease monotonically to below 50\,K with decreasing temperature. With further cooling, $S(T)$ of YCo$_{2}$Zn$_{20}$ passes through a broad peak ($\sim\Theta_{D}$/12 \cite{Jia2008},
where $\Theta_{D}$ is the Debye temperature) expected to be due to phonon-drag \cite{Blatt1976}. On the other hand, $S(T)$ of YFe$_{2}$Zn$_{20}$ shows a local minimum around 14\,K
($\sim\Theta_{D}$/23 \cite{Jia2008}) that is not currently understood. The absolute value of the TEP for YFe$_{2}$Zn$_{20}$ is much smaller than other NFFL systems. A signature of the
spin fluctuation temperature, $T_{sf}$, has been inferred from a shoulder in $A$Fe$_{4}$Sb$_{12}$ ($A$ = Ca, Sr, and Ba) data \cite{Takabatake2006} and as a minimum developed in
$R$Co$_{2}$ ($R$ = Y, Sc, and Lu) data \cite{Gratz2001}. The minimum developed near 14\,K may be related to the signature of spin fluctuation, combined with phonon-drag in the
YFe$_{2}$Zn$_{20}$ system. In the $T\,\rightarrow$\,0\,K limit, the magnitude of $S(T)/T$ of YFe$_{2}$Zn$_{20}$ is larger than that of YCo$_{2}$Zn$_{20}$ as shown in the inset of Fig.
\ref{Zn20YST}.

\begin{figure}
\centering
\includegraphics[width=1\linewidth]{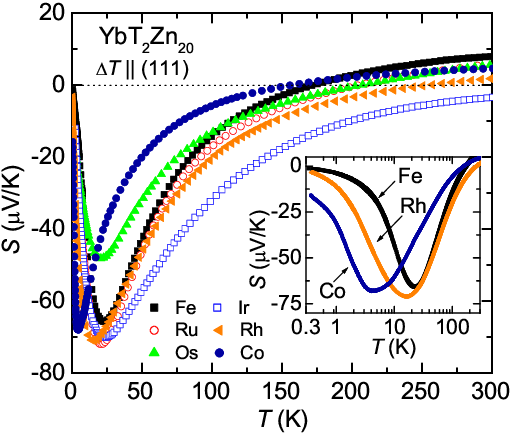}
\caption{Temperature-dependent thermoelectric power, $S(T)$, of YbT$_{2}$Zn$_{20}$ (T = Fe, Ru, Os, Ir, Rh, and Co) in zero applied magnetic field. Inset: $S(T)$ vs. log($T$) for T =
Fe, Rh, and Co.}
\label{Zn20ST1}%
\end{figure}%

The zero field $S(T)$ data of the YbT$_{2}$Zn$_{20}$ (T = Fe, Ru, Os, Ir, Rh, and Co) compounds are plotted in Fig. \ref{Zn20ST1}. In contrast to the isostructural Y-based compounds,
$S(T)$ of the Yb-based compounds exhibits a large, negative minimum (between -75 and -45 \,$\mu$V/K) and the sign of $S(T)$ changing above 150\,K from negative to positive (not
observed in this temperature range for T = Ir). The absolute TEP values of Yb-based compounds are much larger than Y-based compounds at low temperatures, whereas they have a similar
order of magnitude compared to Y-based compounds around 300\,K. A negative, highly enhanced value of the TEP, over the temperature region measured, is typical of those found in other
Yb-based Kondo lattice systems \cite{Foiles1981, Andreica1999, Deppe2008}.

\begin{figure}
\centering
\includegraphics[width=1\linewidth]{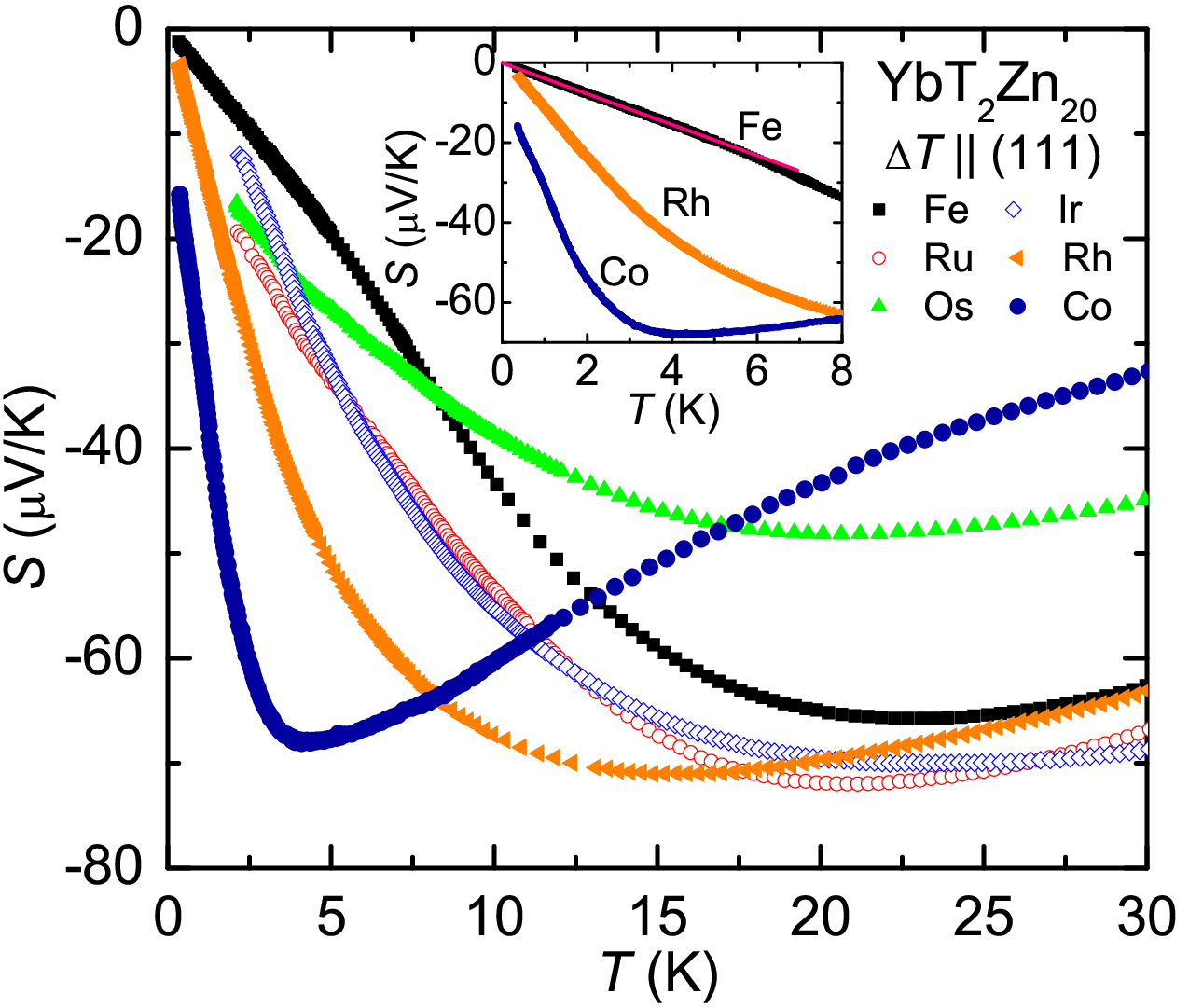}
\caption{Low-temperature $S(T)$ of YbT$_{2}$Zn$_{20}$ compounds in zero applied magnetic field. Inset: $S(T)$ for T = Fe, Rh, and Co below 8\,K. Solid line on the top of the data for
T = Fe is guide to eye.}
\label{Zn20ST2}%
\end{figure}

Figure \ref{Zn20ST2} shows the low temperature $S(T)$ of YbT$_{2}$Zn$_{20}$. For T = Fe and Ru, a broad minimum of $\sim$ -70\,$\mu$V/K is shown at the temperature
$T^{S}_{min}$\,$\sim$\,22\,K. For T = Os, Ir and Rh, a similar broad minimum develops at a temperature of $T^{S}_{min}$\,$\sim$\,16-23\,K, where the width of the peak is wider than
that for T = Fe and Ru. For T = Co, $S(T)$ shows a similar temperature dependence but with the minimum shifted to $T^{S}_{min}$\,$\sim$\,4\,K and it also shows slope changes around
$\sim$\,1\,K and $\sim$\,8\,K. The width of the minimum for T = Co is narrower than that for the other members of this family. Above 10\,K, the absolute value of the TEP for T = Co
reduces more rapidly as the temperature increases than it does for the other YbT$_{2}$Zn$_{20}$ compounds and the sign of the TEP changes from negative to positive close to 150\,K. For
comparison, $S(T)$ curves for T = Co together with T = Fe and Rh are plotted on a semi-logarithmic scale in the inset of Fig. \ref{Zn20ST1}. A smaller local minimum
($\sim$\,-48\,$\mu$V/K) is observed for YbOs$_{2}$Zn$_{20}$. It is not clear at present if this is related to the electrical resistivity measurement that showed a larger residual
resistivity in YbOs$_{2}$Zn$_{20}$ compared to other members (T = Fe, Ru, Ir, and Rh) \cite{Torikachvili2007}. $S(T)$ of YbIr$_{2}$Zn$_{20}$ is negative over the whole temperature
range measured, the sign change from negative to positive being expected around $\sim$\,400\,K, based on a linear extrapolation of $S(T)$ above 250\,K. Below 10\,K (or 3\,K for T =
Co), $S(T)$ data for all compounds show a tendency of approaching zero and reveal linear temperature dependencies that, to varying degrees, approach $S(T)$\,=\,$\alpha T$. The inset
of Fig. \ref{Zn20ST2} shows data down to 0.4 K for T = Fe, Rh, and Co and shows roughly this linear behavior in greater detail.

\begin{figure}
\centering
\includegraphics[width=1\linewidth]{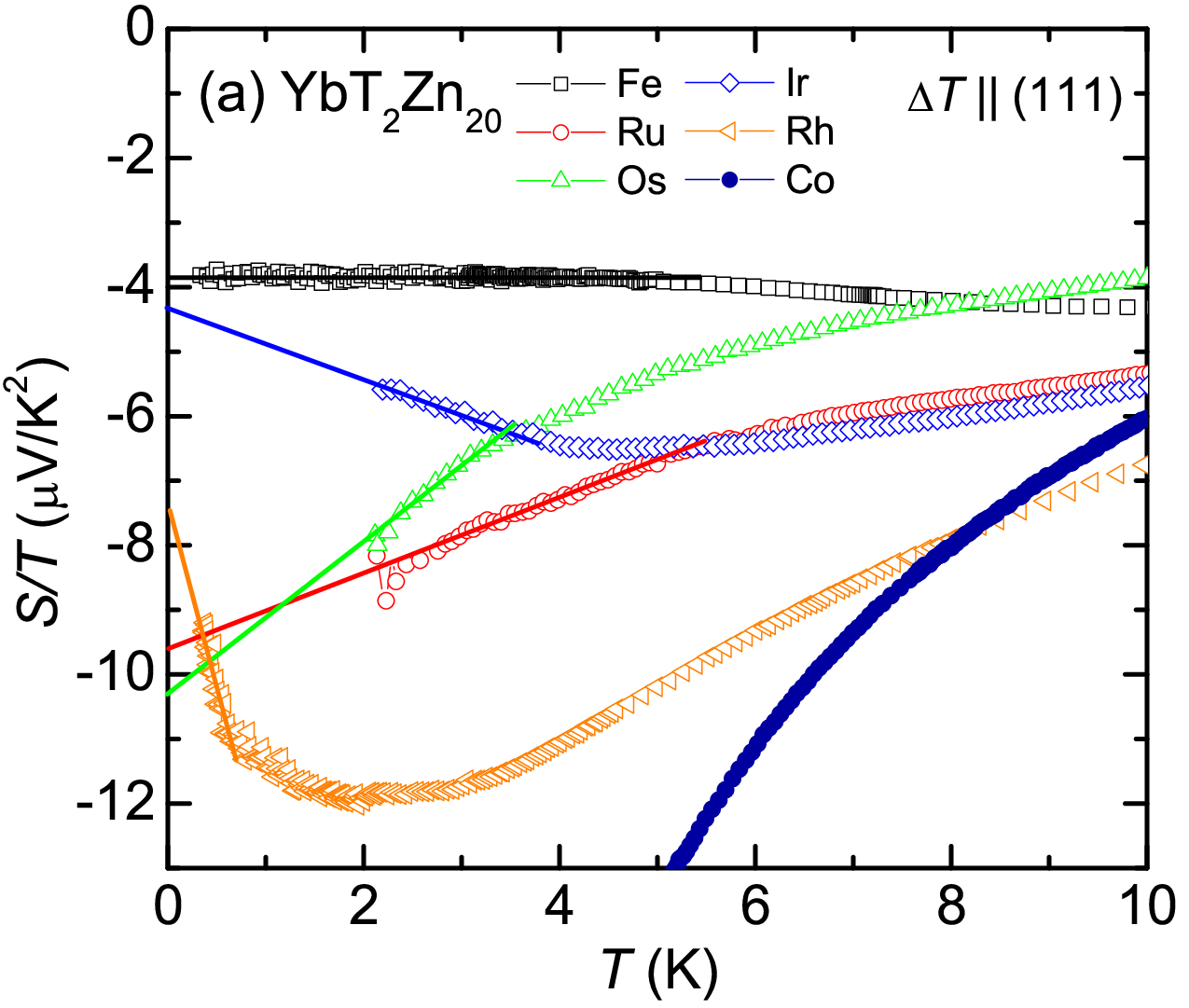}
\includegraphics[width=1\linewidth]{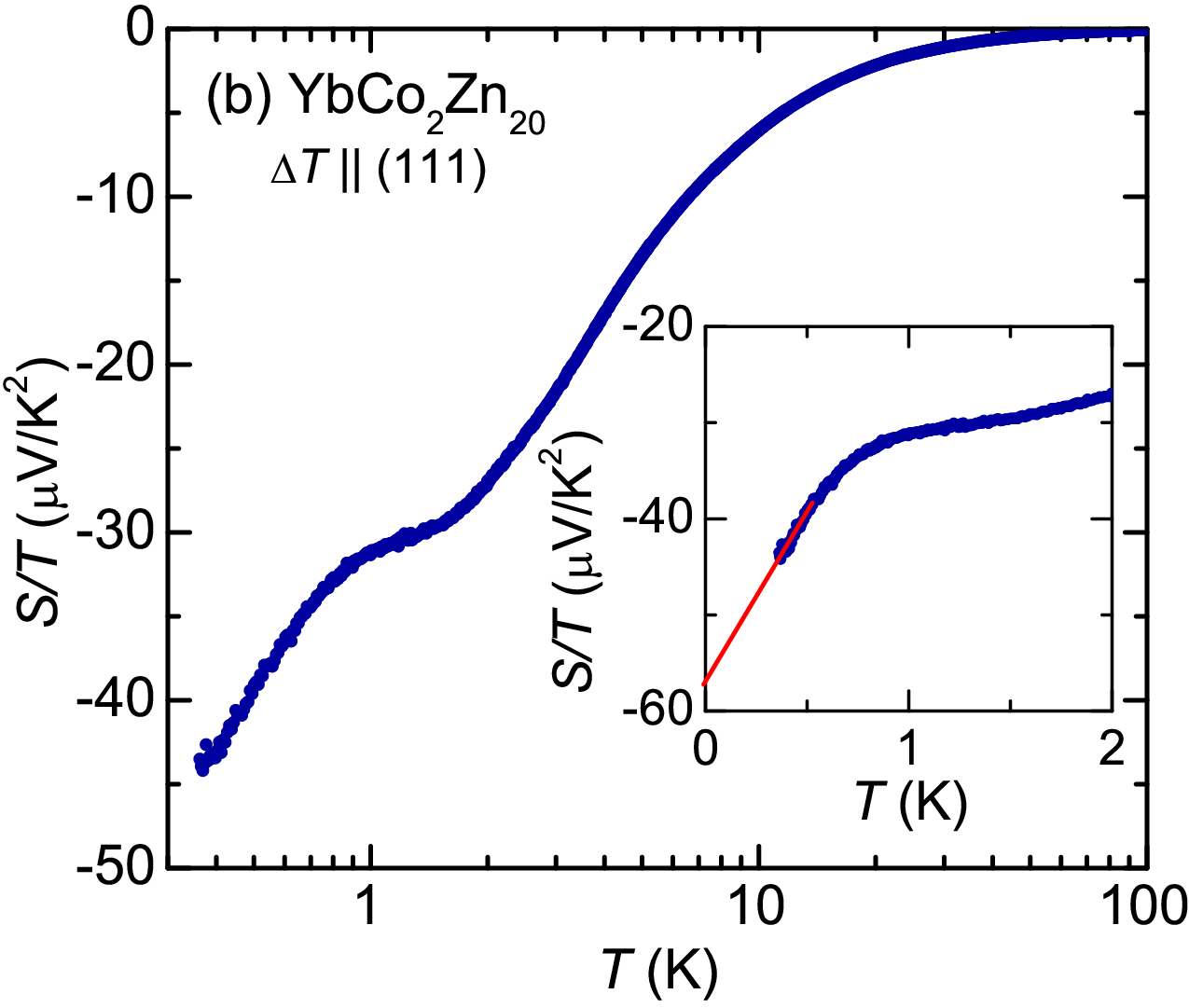}
\caption{(a) $S(T)/T$ vs. $T$ for YbT$_{2}$Zn$_{20}$ below 10\,K in zero applied magnetic field. (b) $S(T)/T$ vs. $T$ for YbCo$_{2}$Zn$_{20}$ below 100\,K. Inset : $S(T)/T$ vs. $T$
below 2 K. The solid lines represent the linear extrapolation curve to $T$ = 0.}
\label{Zn20ST3}%
\end{figure}

The precise temperature dependence of $S(T)$ can be seen more quantitatively in Fig. \ref{Zn20ST3} where the $S(T)/T$ of YbT$_{2}$Zn$_{20}$ below 10\,K is presented. For T = Fe the
clear, linear temperature dependence of TEP, $S(T)/T = \alpha$, is revealed below 4 K. For other compounds, $S(T)/T$ shows an additional temperature dependence (probably due to the CEF
effect and tiny phonon contribution). For T = Rh, the $S(T)/T$ shows a strong temperature dependence at low temperature which is consistent with the temperature dependence of $C(T)/T$
\cite{Torikachvili2007}, where both $S(T)/T$ and $C(T)/T$ revealed a broad peak structure centered around 2$\sim$3 K. For T = Ir, the $S(T)/T$ and $C(T)/T$ also shows a broad feature
around 4 K. Similar linear $S(T)/T$ versus $T$ (i.e. $S \propto T^{2}$) behavior was noted for the majority of HF compounds discussed in Ref. \cite{Behnia2004}. This being said, the
observed temperature dependence of $S(T)/T$ is not similar to the behavior shown in the resistivity, but is consistent with $C(T)/T$. The fact that temperature region manifesting the
Fermi liquid behavior in the resistivity is not the same as that of the $S(T)/T$, can be due to the several additional contribution to $S(T)$. Since the additional contributions
cannot be completely separated, by using a first and simple approximation, the zero temperature limit of $S(T)/T$ has been estimated by linear extrapolating $S(T)$ from 2\,K (or
0.4\,K) to $T$ = 0 (solid lines in Fig. \ref{Zn20ST3}), where the inferred $S(T)/T|_{T\rightarrow 0}$ values for T = Fe, Ru, Os, Ir, and Rh range between -3.8 $\sim$
-10\,$\mu$V/K$^{2}$. For T = Co, the $S(T)/T$ value at 0.4\,K reaches -42\,$\mu$V/K$^{2}$ and is still decreasing (see Fig. \ref{Zn20ST3} (b)). By using a linear extrapolation,
$S(T)/T|_{T\rightarrow 0}$ value for T = Co is found to be $\sim$\,-57\,$\mu$V/K$^{2}$.

\begin{figure}
\includegraphics[width=1\linewidth]{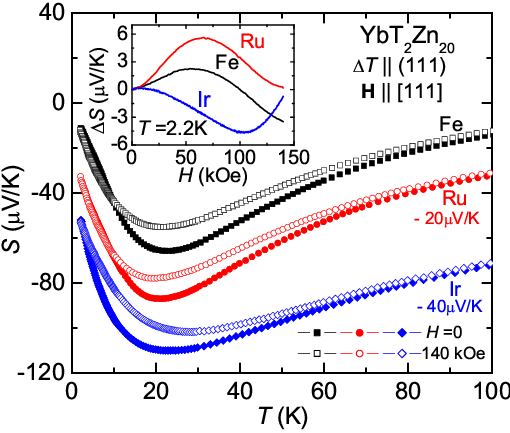}
\caption{$S(T)$ of YbT$_{2}$Zn$_{20}$ (T = Fe, Ru, and Ir) at $H$\,=\,0 (closed symbols) and 140\,kOe (open symbols). For clarity, the data for T = Ru and T = Ir are shifted by -20
$\mu$V/K and -40 $\mu$V/K, respectively. Inset: $\Delta$$S$ = $S(H)-S(0)$ at 2.2 K for T = Fe, Ru, and Ir.}
\label{Zn20SH1}%
\end{figure}

In Fig. \ref{Zn20SH1}, the results of $S(T)$ measurements at $H$ = 0 and 140\,kOe are shown for T = Fe, Ru and Ir. For clarity, the absolute value of the TEP is shifted by
-20\,$\mu$V/K for T = Ru and -40\,$\mu$V/K for T = Ir. A slight change of $T^{S}_{min}$ and a reduction of absolute value are seen for the $H$ = 140 kOe data. Above 100\,K, $S(T)$ for
$H$ = 140\,kOe remains essentially the same as $S(T)$ for $H$ = 0. In the zero temperature limit for $H$ = 140\,kOe data, whereas $S(T)/T$ for T = Ru remain essentially the same,
$S(T)/T$ at 140\,kOe for T = Fe and Ir decrease from $\sim$\,-3.8 to $\sim$\,-6.4\,$\mu$V/K$^{2}$ and from $\sim$\,-4 to $\sim$\,-6.6\,$\mu$V/K$^{2}$, respectively. In the inset, the
TEP measured at $T$ = 2.2\,K is plotted as a function of magnetic field for T = Fe, Ru, and Ir, where $\Delta S = S(H)-S(0)$. An interesting point of this result is the appearance of
a maximum around $\sim$\,70\,kOe for T = Fe and Ru and a minimum around $\sim$\,100\,kOe for T = Ir. For T = Ir the local minimum field shown in TEP is roughly matched with the
metamagnetic-like anomaly seen around $H$ = 120\,kOe in magnetization isotherms, $M(H)$, \cite{Yoshiuchi2009} for \textbf{H}\,$\parallel$\,[110]. For T = Fe and Ru the $M(H)$ data at
$T$ = 2\,K do not show any signature of metamagnetic-like behavior up to 70\,kOe \cite{Mun2010c}, with $M(H)$ being linear in magnetic field for both compounds. In order to clarify
this point, it is necessary to measure $M(H)$ for magnetic fields higher than 70\,kOe, to see whether the anomaly in $S(H)$ is related to features in magnetization or electronic data.

\section{Discussion}

\begin{figure*}
\centering
\includegraphics[width=0.5\linewidth]{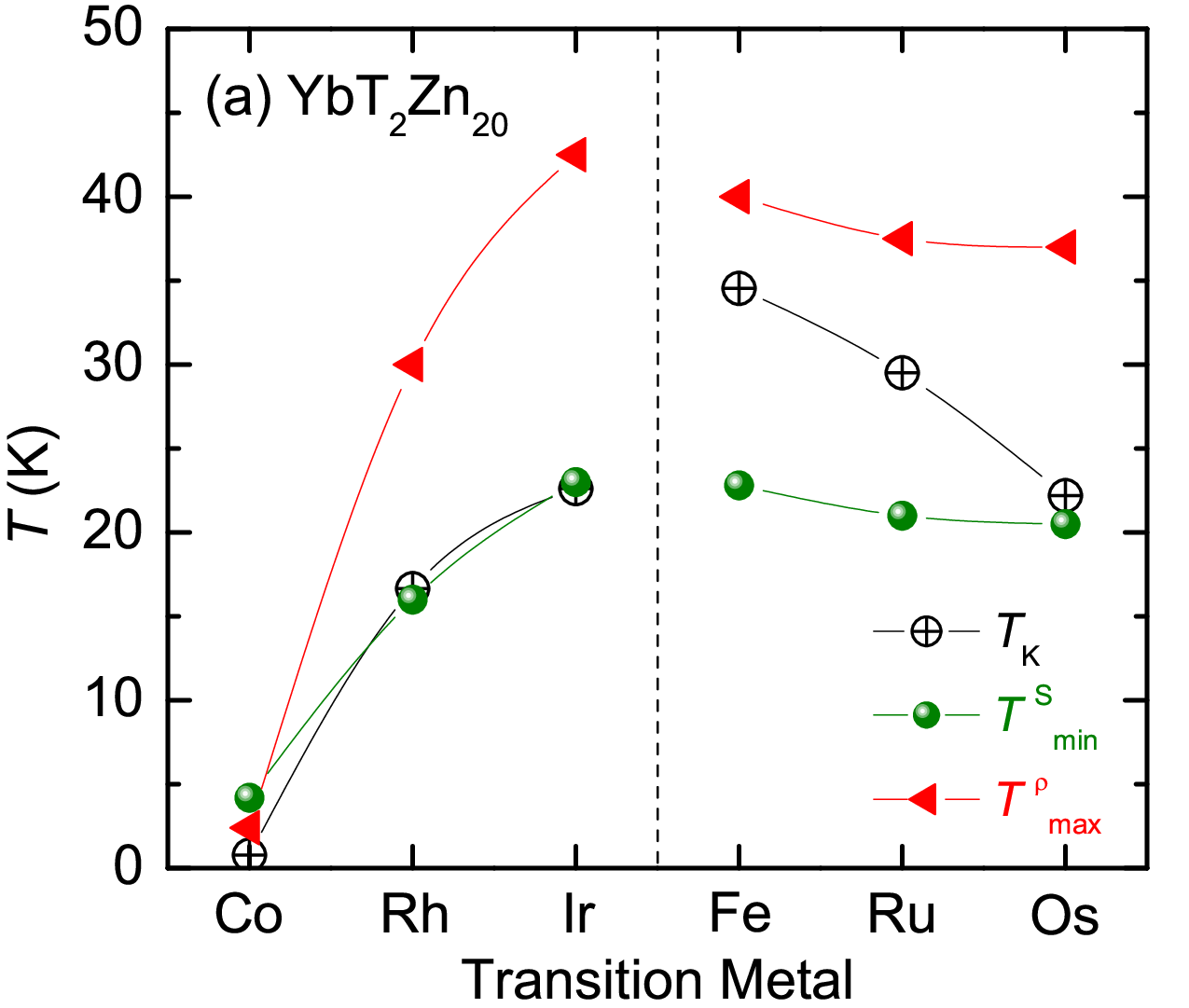}\includegraphics[width=0.5\linewidth]{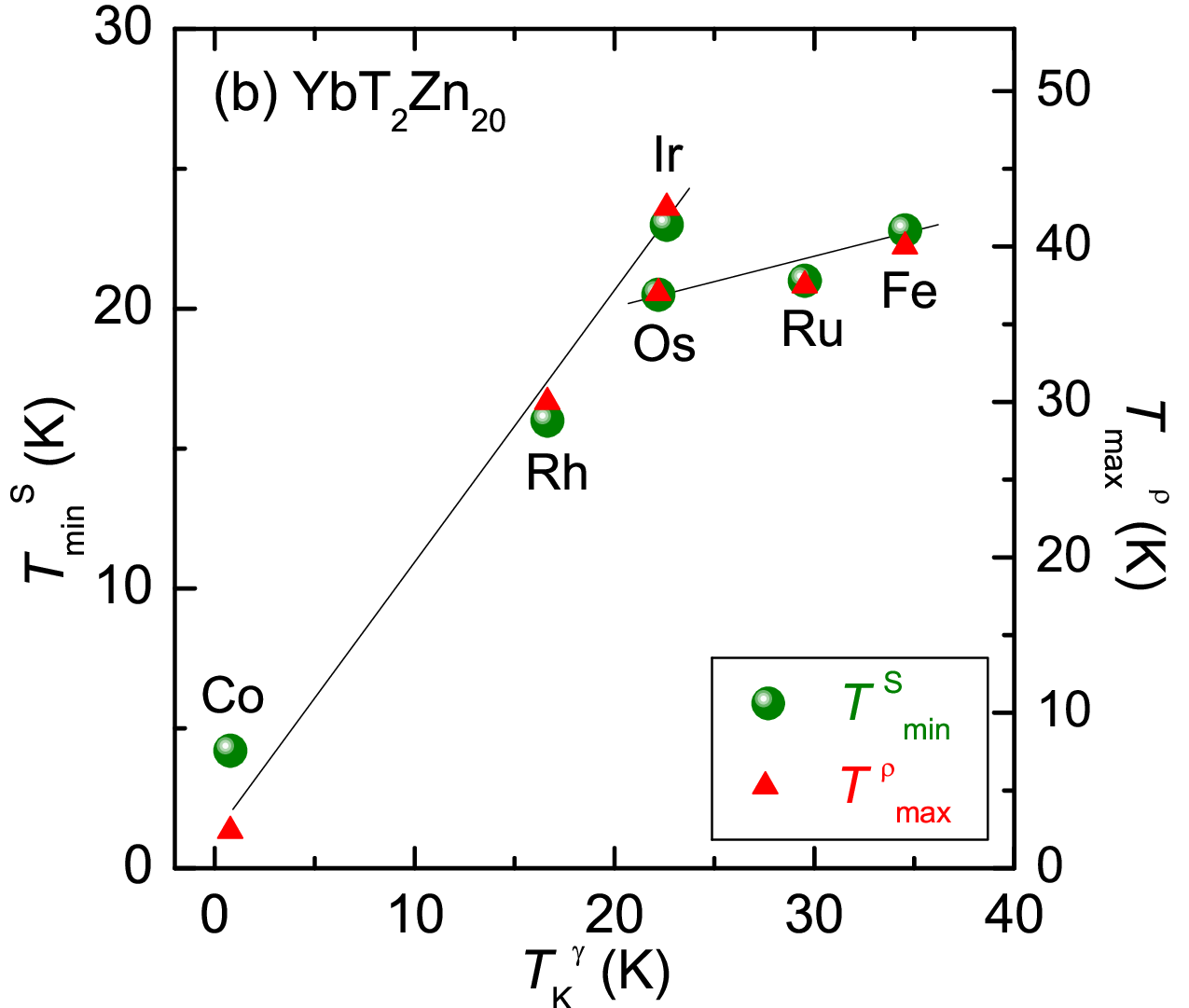}
\caption{(a) Relevant characteristic temperatures in YbT$_{2}$Zn$_{20}$ (T = Fe, Ru, Os, Ir, Rh, and Co). A Kondo temperature ($T_{K}$) calculated from $\gamma$, a local maximum
temperature ($T^{\rho}_{max}$) obtained from the resistivity, and a local minimum temperature ($T^{S}_{min}$) developed in $S(T)$ are plotted as a function of transition metal.
$T^{\gamma}_{K}$ values are taken from Ref. \cite{Torikachvili2007}. Vertical dashed line segregates two columns in the periodic table. (b) Plots of $T^{S}_{min}$ (left) and
$T^{\rho}_{max}$ (right) vs. $T^{\gamma}_{K}$. Solid lines are guide to the eye.}
\label{Zn20Parameters}%
\end{figure*}

Based on earlier thermodynamic and transport measurements of this family \cite{Torikachvili2007}, $S(T)$ data for YbT$_{2}$Zn$_{20}$ (T = Fe, Ru, Os, Ir, Rh, and Co) can be understood
qualitatively by considering the Kondo ($T_{K}$) and CEF ($\Delta/k_{B}$) effects. The compounds in this series appear to be a set of model Kondo lattice systems with varying energy
scales: $T_{K}$ and $\Delta/k_{B}$. In Fig. \ref{Zn20Parameters} (a), the Kondo temperature, $T_{K}$, determined from $\gamma$ \cite{Torikachvili2007} and local minimum temperature,
$T^{S}_{min}$, observed in the zero field $S(T)$ data are plotted as a function of the transition metal, T. The value of $T^{S}_{min}$ correlates strongly with the value of $T_{K}$
for T = Os, Ir, Rh, and Co.

A similar trend can be found in the previously published electrical resistivity, $\rho(T)$, results \cite{Torikachvili2007}. For T = Co, $\rho(T)$ manifests a clear local maximum,
$T^{\rho}_{max}$, around 2.4\,K followed by a logarithmic temperature dependence as temperature decreases. Whereas $T^{\rho}_{max}$ is clear in the $\rho(T)$ data for T = Co,
$\rho(T)$ data from the other members of this family only show a clear local maximum after subtracting the resistivity data of the isostructural LuT$_{2}$Zn$_{20}$ (T = Fe, Ru, Os,
Ir, and Rh) compounds. The local maximum temperatures, $T^{\rho}_{max}$, taken from Ref. \cite{Mun2010a} are plotted in Fig. \ref{Zn20Parameters} (a). The variation of $T^{\rho}_{max}$
follows the same trend as $T^{S}_{min}$ with $T^{\rho}_{max}$ $\sim$ 2$T^{S}_{min}$ even for T = Fe and Ru.

In a Kondo lattice system, a single minimum developed in $S(T)$ is expected when $T_{K}$ is either close to or higher than $\Delta/k_{B}$. Typically, an intermediate valence system
such as YbAl$_{3}$ \cite{Foiles1981} and YbCu$_{2}$Si$_{2}$ \cite{Andreica1999} and a fully degenerate Kondo lattice system such as Yb$_{2}$Pt$_{6}$Al$_{15}$ \cite{Deppe2008} exhibit
a single minimum in the TEP, developing below $T_{K}$. When $T_{K}$\,$<$\,$\Delta/k_{B}$, more than one peak has been frequently observed in the TEP \cite{Andreica1999, Huo2001,
Wilhelm2004, Kohler2008}. The low temperature extremum is usually located close to $T_{K}$, and the high temperature extremum located at 0.4-0.6\,$\Delta/k_{B}$ is attributed to Kondo
scattering off of the thermally populated CEF levels, which is in agreement with theoretical predictions \cite{Bhattacharjee1976, Maekawa1986, Bickers1985, Mahan1997, Zlatic2003,
Zlatic2005}. Therefore, the peak position can represent $T_{K}$ and $\Delta/k_{B}$ as relevant energy scales in Kondo lattice systems.

For the YbT$_{2}$Zn$_{20}$ family, $T_{K}$ and the ground state degeneracy play important roles in the thermodynamic and transport properties. By considering the ground state
degeneracy ($N$\,=\,8 for T = Fe and Ru, and $N$\,=\,4 for T = Os, Rh, Ir, and Co \cite{Torikachvili2007}) it is expected that $T_{K}$\,$\geq$\,$\Delta/k_{B}$ for T = Fe and Ru and
$T_{K} \lesssim \Delta/k_{B}$ for T = Os, Ir, Rh, and Co. Based on this, for T = Fe and Ru, it is reasonable to assume that $T^{S}_{min}$ and $T^{\rho}_{max}$ simply reflect $T_{K}$;
with the fully degenerate case corresponding to $N$\,=\,8. For T = Os, Ir, Rh, and Co, the two extrema in the $S(T)$ data associated with Kondo scattering on the ground state and
thermally populated CEF levels could be expected, however, only one broad peak structure is developed for T = Os, Ir, Rh, and Co. We thus expect that a single broad minimum is
produced by merging more than one peak structure due to the relatively small CEF level splitting ($T_{K}$ $\sim$ $\Delta/k_{B}$).

To reiterate: A strong correlation between the two local extrema $T^{\rho}_{max}$ and $T^{S}_{min}$ develops and remains robust even when dependence on $T_{K}$ appears to break down
(Fig. \ref{Zn20Parameters} (b)). $T^{\rho}_{max}$ $\sim$ 2\,$T^{S}_{min}$ for T = Fe, Ru, Os, Ir, and Rh, and for T = Os, Ir, Rh, and Co $T^{S}_{min}$ $\sim$ $T_{K}$ and
$T^{\rho}_{max}$ $\sim$ 2\,$T_{K}$.

As shown in Fig. \ref{Zn20SH1} the magnetic field dependence of the TEP observed in YbT$_{2}$Zn$_{20}$ (T = Fe, Ru, and Ir) is anomalous. In the simplest case of a two band model, the
carrier density of electrons, $n_{e}$, and holes, $n_{h}$, can be taken as $\frac{1}{2}n$ = $n_{e}$ = $n_{h}$. The diffusion TEP in magnetic field with several assumptions
\cite{Sondheimer1948} can be expressed as
\begin{eqnarray}{\label{Zn20E2}}
\Delta S = S(H) - S(0) = -S(0)\frac{\Upsilon^{2}H^{2} \zeta (1+\zeta)}{1+\Upsilon^{2}H^{2}\zeta^{2}} \nonumber
\end{eqnarray}
where $\Upsilon = 1/nec\rho(0)$, and $\zeta = L_{n}/L_{0}$ with $L_{n} = \frac{1}{3}(\pi k_{B}/e)^{2}$ and $L_{0} = \kappa(0)/\sigma(0)T$ (Lorentz number); $\sigma(0)$ = 1/$\rho(0)$
and $\kappa(0)$ are the electrical conductivity and thermal conductivity, respectively, in zero magnetic field. At low temperatures $L_{0}$ = $L_{n}$, $\rho(0) = \rho_{0}$ (the
residual resistivity), and the diffusion TEP in zero magnetic field is proportional to the temperature, $S(0)$ $\propto$ $T$. Therefore, for simple metals $\Delta S$ = 0 when $T$ = 0,
and for very low temperatures $\Delta S$ $\propto$ $T$. At high temperatures $L_{0}$ = $L_{n}$, and $S(0)$ and $\rho(0)$ are both proportional to temperature, so that $\Delta S$ tends
to zero like 1/$T$ as $T \rightarrow \infty$. In general, the change in the TEP will be too small to be detected at room temperature. Since the magnetoresistance (MR) for T = Fe and
Ru is positive and increases monotonically at 2\,K for \textbf{H}\,$\parallel$\,[111] up to 140\,kOe \cite{Mun2010a}, the change of the TEP ($\Delta S$) should increase or saturate
with increasing magnetic field. The field dependence of the TEP is not consistent with the MR results. Generally, the phonon-drag itself is not sensitive to the applied magnetic field
\cite{Blatt1976}, so it is clear that neither conventional phonon-drag nor diffusion TEP of conduction electrons can account for the magnetic field dependence of the TEP in these
compounds. Thus, multiple factors, such as the Kondo effect and CEF contributions, have to be considered. In order to understand the observed behavior in more detail, a theoretical
analysis of the TEP as a function of field for this systems will be necessary.

\begin{figure*}
\centering
\includegraphics[width=0.5\linewidth]{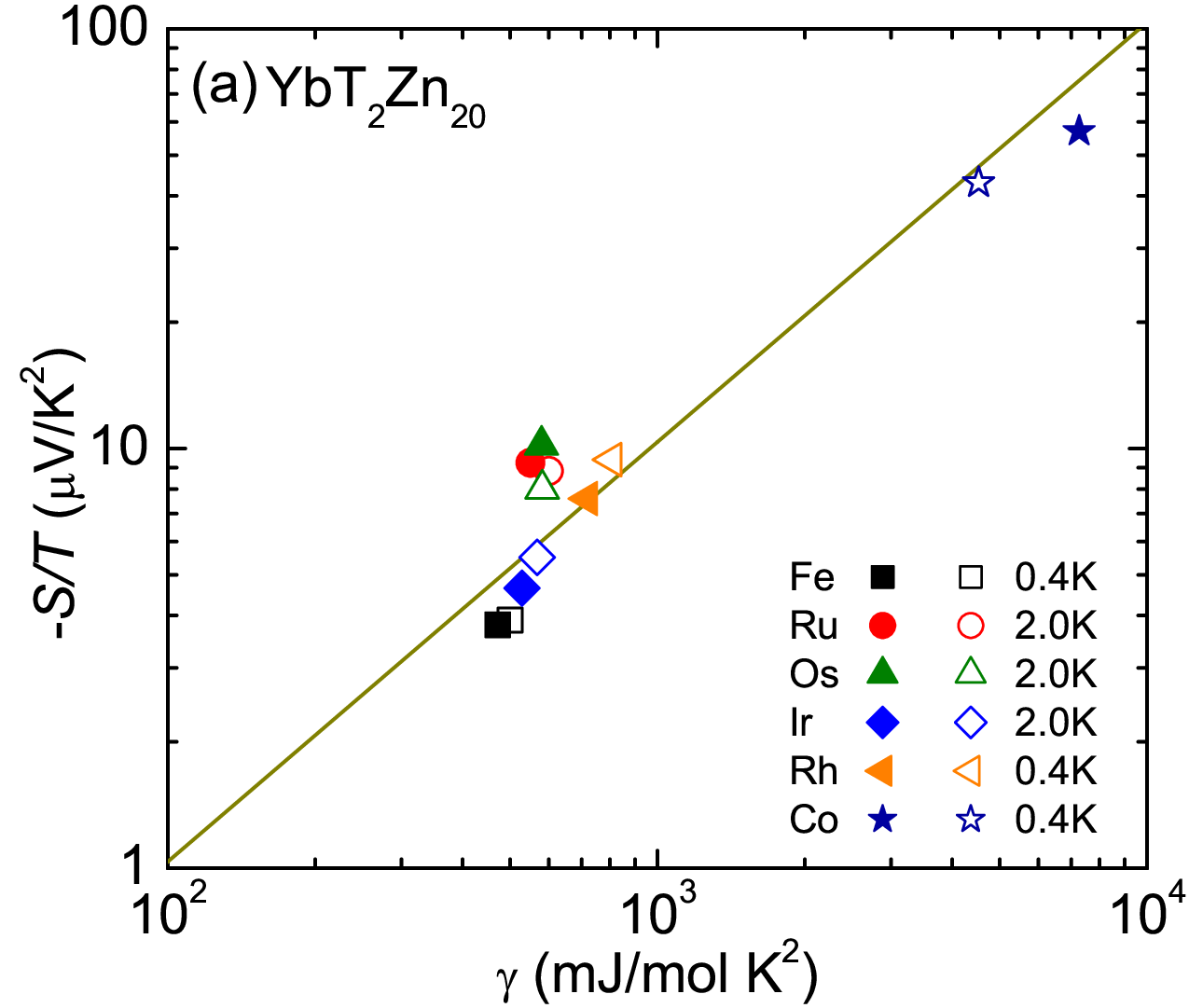}
\includegraphics[width=0.5\linewidth]{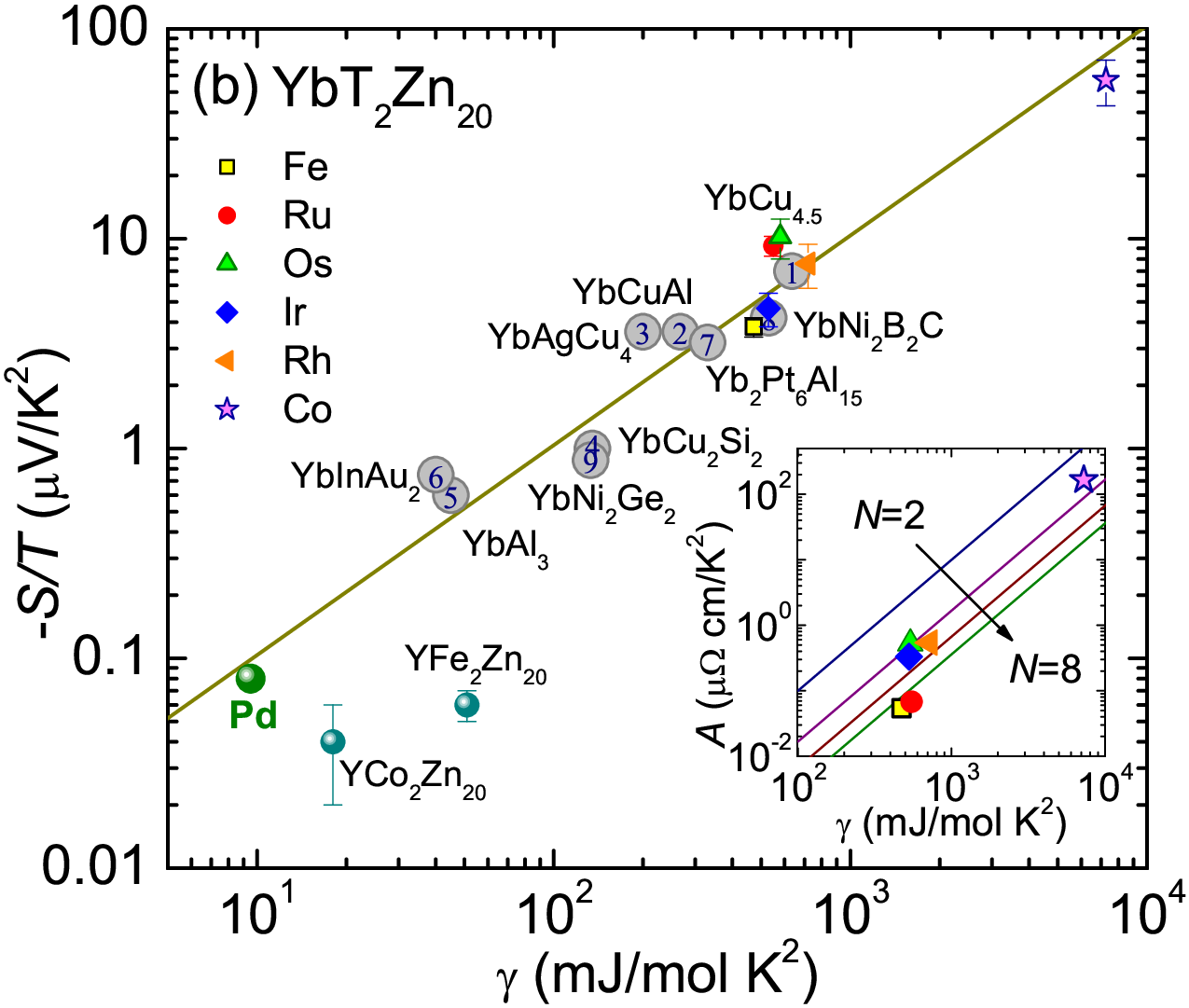}\includegraphics[width=0.5\linewidth]{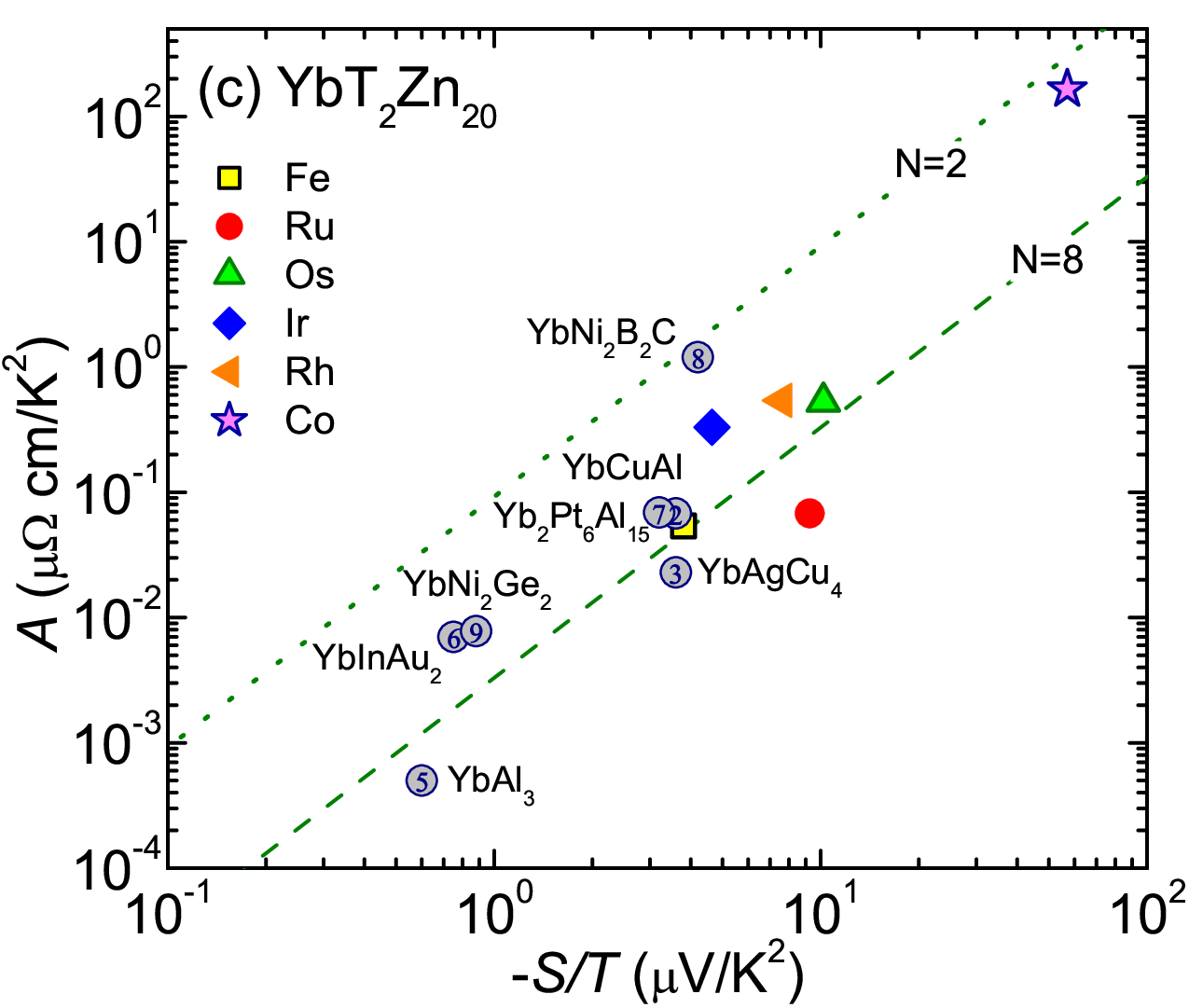}
\caption{(a) -$S(T)/T$ vs. $\gamma$ (log-log) plot of YbT$_{2}$Zn$_{20}$ (T = Fe, Ru, Os, Ir, Rh, and Co). Solid symbols are obtained by the linear extrapolation of
-$S(T)/T|_{T\rightarrow 0}$ and $\gamma \equiv$ $C(T)/T|_{T\rightarrow 0}$. Open symbols are taken by the value of $S(T)/T$ and $C(T)/T$ at base temperature measured (0.4 K for T =
Fe, Rh, and Co and 2 K for T = Ru, Os, and Ir). (b) -$S(T)/T|_{T\rightarrow 0}$ vs. $\gamma$ (log-log) plot of YbT$_{2}$Zn$_{20}$ with reported compounds. The zero temperature limit of
$S(T)/T$ and $\gamma$ for (1) YbCu$_{4.5}$, (2) YbCuAl, (3) YbAgCu$_{4}$, (4) YbCu$_{2}$Si$_{2}$, (5) YbAl$_{3}$, (6) YbInAu$_{2}$, and Pd are taken from the table of Ref.
\cite{Behnia2004}. $S(T)/T$ and $\gamma$ of (7) Yb$_{2}$Pt$_{6}$Al$_{15}$ are taken from Ref. \cite{Deppe2008}. $S(T)/T$ and $\gamma$ of (8) YbNi$_{2}$B$_{2}$C and (9)
YbNi$_{2}$Ge$_{2}$ are taken from Refs. \cite{Behnia2004a, Li2006, Budko1999}, respectively. The solid line represents $\gamma$/$(e N_{A})$. Inset: Kadowaki-Woods plot (log-log plot
of $A$ vs. $\gamma$) of YbT$_{2}$Zn$_{20}$. Symbols are taken from Ref. \cite{Torikachvili2007} and solid lines correspond to $N$ = 2, 4, 6, and 8 based on Ref. \cite{Tsujii2005},
respectively. (c) $A$ vs -$S(T)/T$ (log-log) plot for Yb-based compounds. $S(T)/T$ values are the same as in (b) and $A$ values are taken from Ref. \cite{Tsujii2003, Deppe2008,
Budko1999, Torikachvili2007}. The dotted and dashed line are the $A$ $\sim$ $S/T^{2}$ lines for $N$ = 2 and $N$ = 8, respectively. See details in text.}
\label{Zn20FL}%
\end{figure*}%

Earlier thermodynamic and transport measurements \cite{Torikachvili2007} showed that the $R_{W}$ and K-W ratios of YbT$_{2}$Zn$_{20}$ agree well with the FL picture of the HF ground
state. A clear dependence of the $A/\gamma^{2}$ ratio on the degeneracy $N$ is shown in the inset of Fig. \ref{Zn20FL} (b), where the $A$ and $\gamma$ values are taken from Ref.
\cite{Torikachvili2007} and lines for degeneracies $N$ are based on Ref. \cite{Tsujii2005}. A Fermi liquid state can also be characterized by the ratio between $\gamma$ and the zero
temperature limit of $S(T)/T$ \cite{Behnia2004, Grenzebach2006, Zlatic2007}: a ``quasi universal" ratio $q$ = $N_{A} e S/\gamma T$ remains close to $q$ = $\pm$\,1 for metals and the
sign of $q$ depends on the type of carriers. Although for strongly correlated electronic materials like HF systems, a single band and single scattering process is not generally
thought to be sufficient for explaining the strong correlation effects, given that $C(T)/T$ and $S(T)/T$ are most sensitive to the position of the heavy band, a quasi universal ratio
is expected to hold at low temperatures \cite{Miyake2005, Kontani2003}.

As shown in Figs \ref{Zn20YST} and \ref{Zn20ST3}, a clear Fermi liquid behavior, $S(T)/T = \alpha$, is shown for YFe$_{2}$Zn$_{20}$ and YbFe$_{2}$Zn$_{20}$. The other compounds in this
family reveal a deviation from the linear temperature dependence of TEP within the measured temperature range, where the feature shown in temperature dependence $S(T)/T$ is similar to
that of $C(T)/T$. Since several effects in $S(T)/T$ are included, such as CEF and phonon contribution, $S(T)/T$ vs $C(T)/T$ is compared both at finite temperature and the zero
temperature limit (the latter being the same approach used for the majority of HF systems shown in Fig. 1 of Ref. \cite{Behnia2004}). In Fig. \ref{Zn20FL} (a), $S(T)/T$ at lowest
temperature measured vs $C(T)/T$ is plotted, where 0.4 K for T = Fe, Rh, Co and 2 K for T = Ru, Os, Ir, as shown by open symbols. The zero temperature limit of $S(T)/T$, estimated by
simple linear extrapolation of $S(T)/T$ from 2 K (or 0.4 K) to $T$ = 0 (solid lines in Fig. \ref{Zn20ST3} (a)), is also plotted (closed symbols in Fig. \ref{Zn20FL} (a)). Both the
zero temperature limit and $S(T)/T$ value at the lowest temperature measured are locating close to the line with $q$ = -1.

The experimental correlation between the zero temperature limit of $S(T)/T$ and $\gamma$ for YbT$_{2}$Zn$_{20}$ (T = Fe, Ru, Os, Ir, Rh, and Co) and YT$_{2}$Zn$_{20}$ (T = Fe and Co)
is presented in Fig. \ref{Zn20FL} (b). Though error bars cannot be determined exactly from the present data because additional contributions such as phonon and CEF effect in $S(T)/T$
cannot be completely separated, one can use the difference between the base temperature value measured and the extrapolated value to $T$ = 0. The error bars in Fig. \ref{Zn20FL} (b)
are based on the difference between the values determined from the base temperature value measured and linear extrapolation to $T$ = 0 (other experimental error has not been applied to
the error bar in Fig. \ref{Zn20FL} (b), where the experimental error is smaller than the error bar determined). Given that Fig. \ref{Zn20FL} (b) is a log-log plot, spanning orders of
magnitude, the error bars are of limited concern. For comparison, data for several other Yb-based HF compounds as well as Pd are also plotted in the same figure \cite{Behnia2004a}. The
calculated $q$ values of Yb-based compounds vary from -0.77 for T = Fe to -1.4 for T = Rh, which are close to the value $q$ = -1, expected for hole-like charge carriers.

As shown in Fig. \ref{Zn20FL} (b), each Yb-based data point is close to a line represented by $q$ = -1 which means that the zero temperature limit of $S(T)/T$ is strongly correlated
to $\gamma$ due to the enhanced density of state at the Fermi level; the larger density of states at the Fermi level results in a larger $\gamma$ and $S(T)/T|_{T\rightarrow 0}$. For
YFe$_{2}$Zn$_{20}$ and YCo$_{2}$Zn$_{20}$, though, the calculated $q$ ($\sim$ -0.1) value is 10 times smaller than that for the free electron case. This can be understood, at least in
part, by appreciating the fact that, since the $\gamma$ term is not dominated by a huge Yb-contribution, the $\gamma$ value should probably be expressed in terms of per-mole-atomic,
reduced by a factor of $\sim$ 20. Such a reduction of $\gamma$ would place these two data points more or less on the $q$ = -1 line.

Given the large range of $S(T)/T$ and $\gamma$ values found for the YbT$_{2}$Zn$_{20}$ compounds we can examine the direct correlation found between $S(T)/T$ and $A$ for Yb-based
materials. In Fig. \ref{Zn20FL} (c) data from Ref. \cite{Tsujii2003, Deppe2008} along with our data for the six YbT$_{2}$Zn$_{20}$ compounds are shown. At the grossest level, larger
$S(T)/T$ values corresponds to larger $A$ values. More quantitatively, in a naive picture, since $A \propto \gamma^2$ and $S(T)/T \propto \gamma$, then $A \propto (S(T)/T)^2$. In Fig.
\ref{Zn20FL} (c), the dotted and dashed lines are the $A$ $\sim$ $(S/T)^{2}$ lines with prefactors appropriate for $N$ = 2 and $N$ = 8, respectively, discussed for the generalized K-W
ratio; $A/(S/T)^2$ = 9.216 $\times$ 10$^{4}$ /$(0.5N(N-1)$ $\Omega$ cm/K$^{2}$/(C/mol $\cdot$ V/K)$^2$, where $A/\gamma^{2}$ = 1.0$\times$10$^{-5}$ / $(0.5N(N-1)$ $\Omega$
cm/(J/mol$\cdot$K)$^{2}$ \cite{Tsujii2005} and $q$ = -1 \cite{Behnia2004} are used. As can be seen, virtually all of the data, for a diverse set of structures, Kondo temperatures, and
Yb-content, fall between these two extremes, simultaneously (i) giving some sense that they are capturing the salient correlation between these values and (ii) demanding a more
detailed and formal, theoretical examination of the relation between these two transport properties in strongly correlated electron systems.

\section{Summary}
The thermoelectric power measurements on the YbT$_{2}$Zn$_{20}$ (T = Fe, Ru, Os, Ir, Rh, and Co) compounds are in agreement with the behavior observed in many heavy fermion Kondo
lattice systems. The evolution of the local minimum in $S(T)$ and the local maximum (coherence temperature) in $\rho(T)$ with variation of the transition metals can be understood
based on the energy scale of Kondo temperature in conjunction with the influence of the crystalline electric field splitting. The large value of $S(T)/T$ in the zero temperature limit
can be scaled with the electronic specific heat coefficient, $\gamma$, which is reflected by a strong correlation via the universal ratio $q = N_{A} e S/\gamma T$ and confirms the
validity of Fermi-liquid descriptions. In addition, for a wide range of Yb-based materials there is a clear, apparently simple, correlation between $S(T)/T$ and $A$.

\begin{acknowledgments}
This work was supported by the U.S. Department of Energy, Office of Basic Energy Science, Division of Materials Sciences and Engineering. The research was performed at the Ames
Laboratory. Ames Laboratory is operated for the U.S. Department of Energy by Iowa State University under Contract No. DE-AC02-07CH11358.
\end{acknowledgments}

\end{document}